%% file: RGA-2019-01-27.tex
\documentclass[conference]{IEEEtran}
\IEEEoverridecommandlockouts
\usepackage{amsthm, amsfonts}
\usepackage{graphicx}
\usepackage{amsmath}
\usepackage{mathtools}
\usepackage{enumitem}

\usepackage[ruled,linesnumbered,resetcount]{algorithm2e}
\usepackage{algpseudocode}
\usepackage{booktabs}
\usepackage[table]{xcolor}
\usepackage{pdflscape}
\usepackage{longtable}
\usepackage{multirow}
\usepackage{threeparttablex}
\usepackage{graphicx}
\usepackage[compatibility=false]{caption}
\usepackage[labelformat=simple]{subcaption}
\usepackage[numbers]{natbib}
\usepackage{url}
\usepackage{rotating}

\usepackage{hyperref}
\usepackage{lipsum}
\usepackage{import}
\usepackage[acronym]{glossaries}
\makeglossaries

\usepackage{graphicx}
\usepackage{amsmath}
\usepackage{mathtools}
\usepackage[ruled,linesnumbered,resetcount]{algorithm2e}

\usepackage{pdfpages}
\usepackage{import}
\usepackage{multicol}
\usepackage{lineno}
\usepackage{tabularx}

\hypersetup{
	bookmarks=false,
	colorlinks=true,       % false: boxed links; true: colored links
	linkcolor=blue,          % color of internal links (change box color with linkbordercolor)
	citecolor=green,        % color of links to bibliography
	filecolor=magenta,      % color of file links
	urlcolor=cyan         
}

\loadglsentries{Glossaries}

\newcommand{\scalefigure}{0.20}

\def\BibTeX{{\rm B\kern-.05em{\sc i\kern-.025em b}\kern-.08em
    T\kern-.1667em\lower.7ex\hbox{E}\kern-.125emX}}
\begin{document}

\title{A Heuristic Based on Randomized Greedy Algorithms for the Clustered Shortest-Path Tree Problem\\
}

\author{
	\IEEEauthorblockN{
		Huynh Thi Thanh Binh\IEEEauthorrefmark{2}
		Pham Dinh Thanh\IEEEauthorrefmark{1},
		Do Dinh Dac\IEEEauthorrefmark{2},
		Nguyen Binh Long\IEEEauthorrefmark{2},
		Le Minh Hai Phong\IEEEauthorrefmark{2}}
	\IEEEauthorblockA{
		\IEEEauthorrefmark{1} Faculty of Mathematics - Physics - Informatic, Taybac University, Vietnam \\
		Email: thanhpd05@gmail.com \\		
		\IEEEauthorrefmark{2} School of Information and Communication Technology, Hanoi University of Science and Technology, Vietnam\\
		Email: binhht@soict.hust.edu.vn, $\left\{\text{dacdodinh99, binhlongmm99, aquariuslee99}\right\}$@gmail.com\\		
	}
}

\maketitle

\begin{abstract}
\glspl{rga} are interesting approaches to solve problems whose structures are not well understood as well as problems in combinatorial optimization which incorporate the random processes and the greedy algorithms. This paper introduces a new algorithm  that combines the major features of \glspl{rga} and \gls{spta} to deal with the \gls{clustp}. In our algorithm, \gls{spta} is used to determine the shortest path tree in each cluster while the combination between characteristics of the \glspl{rga} and search strategy of  \gls{spta} is used to constructed the edges connecting clusters. To evaluate the performance of the proposed algorithm, Euclidean benchmarks are selected. The experimental investigations show the strengths of the proposed algorithm in comparison with some existing algorithms. We also analyze the influence of the parameters on the performance of the algorithm.
\end{abstract}


%%================------------------Sec_Introduction-------------------=============
\section{Introduction}
\label{Sec_Introduction}
\input{Sections/Sec_Introduction}

%================------------------Sec_Notation_and_definitions-------------------=============
\section{Notation and definitions}
\label{Notation_and_definitions}
\input{Sections/Sec_Notation_and_definitions}

%%================------------------Sec_Related_Works-------------------=============

\section{Related works}
\label{Sec_Related_Works}
\input{Sections/Sec_Related_works}

%%================------------------Sec_Proposed_Algorithm-------------------=============
\section{Proposed Algorithm}
\label{Sec_Proposed_Algorithm}
\input{Sections/Sec_Proposed_Algorithm}

%%================------------------Sec_Computational_results-------------------=============
\section{Computational results}
\label{Sec_Computational_results}
\input{Sections/Sec_Computational_Results}

%%%------------------------------------Sec_Conclusion---------------------------------------------%%%
\section{Conclusion}
\label{Sec_Conclusion}
\input{Sections/Sec_Conclusion}

%%%------------------------------------acknowledgements---------------------------------------------%%%
%\section*{Acknowledgment} 
%\label{Acknowledgment}
%\input{Sec_Acknowledgment}	

\bibliographystyle{IEEEtran}
\bibliography{references}   % name your BibTeX data base

\onecolumn
\begin{landscape}
	\input{Table_Data/GammaType1}
\end{landscape}

%\clearpage
%\onecolumn
%\begin{landscape}
%	
\input{Table_Data/ResultsType1}	
%	
%	\input{Table_Data/ResultsType34}	
%	
\input{Table_Data/ResultsType5}	
%	
\input{Table_Data/ResultsType6}	
%\end{landscape}

%\vspace{12pt}
%\color{red}
%IEEE conference templates contain guidance text for composing and formatting conference papers. Please ensure that all template text is removed from your conference paper prior to submission to the conference. Failure to remove the template text from your paper may result in your paper not being published.

\end{document}

%% file: Glossaries.tex
\newacronym{ga}{GA}{Genetic Algorithm}
\newacronym{ea}{EA}{Evolutionary Algorithm}
\newacronym{mfo}{MFO}{Multifactorial Optimization}
\newacronym{mfea}{MFEA}{Multifactorial Evolutionary Algorithm}
\newacronym{clustp}{CluSPT}{Clustered Shortest-Path Tree Problem}
\newacronym{sto}{STO}{single-tasking of Genetic Algorithm}
\newacronym{clumrct}{CluMRCT}{Minimum Routing Cost Clustered Tree Problem}
\newacronym{clutsp}{CluTSP}{Clustered Traveling Salesman Problem}
\newacronym{ftsp}{FTSP}{Family Traveling Salesman Problem}
\newacronym{gtsp}{GTSP}{Generalized Traveling Salesman Problem}
\newacronym{gp}{GP}{Genetic Programming}
\newacronym{ant}{ACO}{Ant Colony Optimization}
\newacronym{grasp}{GRASP}{Greedy Randomized Adaptive Search Procedure}

\newacronym{tsp}{TSP}{Traveling Salesman Problem}
\newacronym{steinertp}{STP}{Steiner Tree Problem}
\newacronym{clusteinertp}{CluSteinerTP}{Clustered Steiner Tree Problem}
\newacronym{uss}{USS}{Unified Search Space}
\newacronym{de}{DE}{Differential Evolution Algorithm}
\newacronym{pso}{PSO}{Particles Warm Optimization Algorithm}
\newacronym{rmp}{rmp}{Probability of Random Mating}
\newacronym{vrp}{VRP}{Vehicle Routing Problem}
\newacronym{cvrp}{CVRP}{Capacitated Vehicle Routing Problem}

%----------------------This Paper--------------------------
\newacronym{ltga}{LTGA}{Linkage Tree Genetic Algorithm}
\newacronym{mfed}{MFEA}{Multifactorial evolutionary algorithm}
\newacronym{mfltga}{MF-LTGA}{Multifactorial Linkage Tree Genetic Algorithm}

\newacronym{nrga}{HBRGA}{Heuristic Based on Randomized Greedy Algorithm}
\newacronym{rga}{RGA}{Randomized Greedy Algorithm}
\newacronym{spta}{SPTA}{Shortest Path Tree Algorithm}

%% file: Sections/Sec_Introduction.tex
The incorporation of randomness in many algorithms can be crucial to solve various problems in the fields of computer sciences. For example, one of the most well-known algorithms in randomized search heuristics~\cite{li_heuristic_1998,vose_random_1999} is \gls{ea}~\cite{neumann_randomized_2007} which is successfully applied to deal with numerous type of problems. In \gls{ea}~\cite{agoston_eiben_2003, back_evolutionary_1996}, the randomness appears in most of key components of the algorithms such as  the process of population initialization, the crossover and mutation operators. Another remarkable algorithm is \gls{ant}~\cite{dorigo_ant_2019} which currently  also attracts a lot of attention in the scientific community to extend and apply to many hard optimization problems. In \gls{ant}, the randomness is used to construct ant solutions~\cite{dorigo_ant_2019}.

Recently, some variants of heuristic search algorithms are proposed such as \gls{grasp}~\cite{marquez_overview_2018, resende_grasp_2014}, \glspl{rga}~\cite{gao_randomized_2018,poloczek_randomized_2012,zhang_greed_2014}, etc. in which \glspl{rga} are notable approaches because of the fact that \glspl{rga} takes both advantages of the Greedy Algorithm (fast approach for finding a solution, and implementation in a very simple way~\cite{gao_randomized_2018}) and random search methods (simple implementation, easy computation; effective when constraints are involved or the gradient is burdensome~\cite{li_heuristic_1998}). \glspl{rga} randomizes the greedy step to prevent the algorithm from producing bad solutions. A major difference in comparison with deterministic greedy algorithm is that \glspl{rga} selects a solution component from a set of available components at each step according to the given probability distribution. Therefore, \glspl{rga} use an additional parameter ($\gamma$) to compute the amount of greediness. \glspl{rga} has been proved to be effective when solving some problems~\cite{gao_randomized_2018,poloczek_randomized_2012,zhang_greed_2014}.

Nowadays, the problem of communication network connection arises as an urgent demand. As a result, clustered tree problems such as the Clustered Steiner Tree problem, the Inter Cluster Tree Problem and the \gls{clustp} also attract a lot of interests for their wide range of applications. In particular, the \gls{clustp} has some applications such as: goods distribution, water supplies, and fiber optic cable network. 

There have been some studies to deal with the \gls{clustp}. Those studies are based on the \gls{mfea} and focused on searching solutions of many problems simultaneously. Although those algorithms have been proven to be effective in some test cases, in the aspect of producing a good solution that approximates the global optimum within a small amount of time, the existing algorithms have revealed some restrictions.

As a consequence, we propose a random heuristic search for the \gls{clustp}. The proposed algorithm is a combination between \glspl{rga} and \gls{spta}. The main objective of the \gls{spta} is to construct the shortest-path tree for each cluster while the objective of \glspl{rga} is to build the edges connecting the clusters. To select an edge to connect a visited cluster to an unvisited cluster, our algorithm, \gls{nrga}, defines a reward of an edge based on the weight of the edge and the cost of the shortest-path tree of the unvisited cluster rooted at one end of the edge. Then, \gls{nrga} selects an edge according to a probabilistic distribution formula which is computed based on the rewards of the edges. Extensive computational results are reported and discussed for various Euclidean benchmarks. The obtained results demonstrate the efficiency of our approach compared to some existing methods.

The major contributions of this work as following:
\begin{itemize}
	\item Propose a mechanism for combining the \glspl{rga} with the \gls{spta} to deal with the \gls{clustp}.
	\item Suggest a strategy for evaluating rewards of edges.
	\item Introduce a random effect during the edge selection to construct temporary solution.
	\item Analyze the effects of parameters on the performance of the proposed algorithm.
	\item Experiment to evaluate the efficiency of the proposed algorithm.
\end{itemize}

This paper is organized as follows. Section~\ref{Sec_Related_Works} introduces related works. Section~\ref{Sec_Proposed_Algorithm} describes the proposed algorithm. Section~\ref{Sec_Computational_results} presents and discusses experimental results. The paper concludes in section~\ref{Sec_Conclusion} with discussions on the future extension of this research.

%% file: Sections/Sec_Notation_and_definitions.tex
In this paper, we consider simple connected undirected graphs. For a graph $G = (V, E, w)$, $V$ is the vertex set, $E$ is the edge set, and $w$ is the nonnegative edge length function of the graph. The weight of an edge (u, v) is denoted by $w(u, v)$.

For a graph $G$, $V(G)$ and $E(G)$ denote the vertex and the edge sets, respectively. For a vertex subset $U$, the sub-graph of $G$ induced by $U$ is denoted by $G[U]$. For a vertex set $V$, a collection $\{V_i | 1 \leq i \leq k\}$ of subsets of $V$ is a partition of $V$ if the subsets are mutually disjoint and their union is exactly $V$. A path of $G$ is simple if no vertex appears more than once on the path. In this paper we consider only simple paths.

$Q$ is the set of all the clusters that have not been included in the current solution. At the beginning, $Q$ contains all k clusters.

Root cluster is the cluster that has the source vertex.

Local root of a cluster is the first vertex on the path from the source vertex to a vertex in that cluster. Let $root[i]$ be the local root of cluster $V_i$.

The distance between the root cluster and another cluster is defined by the total cost of the shortest path from the source vertex to the local root of that cluster. Let $dis[i]$ be the distance from the root cluster to cluster $V_i (i=1, \ldots, k)$, so we first initialize:
\begin{itemize}
	\item $dis[1] = 0$,
	\item $dis[i] = +\infty$, $i \in {2,\ldots, k}$
\end{itemize}

For each cluster $V_i$, $d(root[i], v)$ is the distance from the local root of cluster $V_i$ to a vertex $v \in V_i$.

For a given spanning tree $T$ of $G = (V, E, w)$ and $u, v \in V$, let $d_T(u, v)$ denote the shortest path length between $u$ and $v$ on $T$. 

The \gls{clustp} is defined as following:
\begin{center}
	\begin{tabular}{l p{6.5cm}} %<----2 cot
%	\begin{tabular}{l p{12cm}}
%		\hline 
%		\multicolumn{2}{c}{\textbf{Clustered Shortest-Path Tree Problem}} \\ 
%		\hline 
		\hline 
		\textbf{Input}:		
		&  - A weighted undirected graph $G = (V, E, w)$.\\
		&  - Vertex set $V$ is partitioned into $k$ clusters ${V_1, V_2,\ldots, V_k}$.\\
		&  - A source vertex $s \in V$.\\
		\hline
		\textbf{Output}:   	
		&  - A spanning tree $T$ of $G$.\\
		&  - Sub-graph $T[V_i] (i = 1,\ldots, k)$ is a connected graph.\\
		\hline 
		\textbf{Objective}: & $\displaystyle \sum_{v \in V} d_{T}(s,v) \rightarrow $ min\\
%		& where $d_{T}(u, v)$ is the cost of shortest path from vertex $u$ to vertex $v$ on $T$.\\ 
		\hline 
	\end{tabular}
\end{center}

%% file: Sections/Sec_Related_works.tex
Randomized search heuristics are successfully applied to various types of problems, especially for the problems having  not well-understood structure or problems in combinatorial optimization~\cite{neumann_randomized_2007}, such as \gls{ea} for \gls{tsp}~\cite{nagata_new_2006} and \gls{cvrp}~\cite{nagata_edge_2007}; simulated annealing for \gls{cvrp}~\cite{wei2018simulated}; \gls{grasp} for quadratic assignment problem~\cite{mateus_grasp_2011}, for Steiner problem~\cite{ribeiro_hybrid_2002} and for generalized minimum spanning tree problem~\cite{ferreira_grasp-based_2012}; \glspl{rga} for Covering Problems~\cite{gao_randomized_2018} and Maximum Matching Problem~\cite{poloczek_randomized_2012,aronson_randomized_1995}; etc. Each algorithm has its own advantages but in the aspect of looking for good approximations of optimal solutions within a small amount of time and easy to implement, the \glspl{rga} are one of the most suitable approaches.

The conception of \glspl{rga} is inspired from incorporating randomness into greedy algorithm. In \glspl{rga}, randomness is used in greedy selection steps~\cite{gao_randomized_2018} for improving the quality of solutions. \glspl{rga} proves to be effective in solving various types of problems.

Yuan Zhang et.al.~\cite{zhang_greed_2014} introduced a randomized greedy algorithm for dependency parsing. The main benefit of proposed algorithm is to tie to the number of local maxima in the search space. The authors point out that a randomized greedy method of inference surpasses the state-of-art performance in dependency parsing.

In~\cite{dyer_randomized_1991}, a randomized version of the greedy algorithm is presented to determine a large matching in a graph. The authors point out that on some classes of sparse graphs, \glspl{rga} performs significantly better than the worst-case and the ratio of the expected size of the randomized greedy matching to maximum size of matching is at least 0.7690. J.~Aronson~et.al.~\cite{aronson_randomized_1995} considered the modified randomized greedy algorithm for finding a matching M in an arbitrary graph. The authors proved that the maximum matching within a factor of at leats $\frac{1}{2}+\frac{1}{400,000}$. The authors also received stronger results for tree and sparse graphs. Mathhias~Poloczek~et.al.~\cite{poloczek_randomized_2012} invented the Constrast Analysis technique to analyze the lower bound of the modified randomized greedy algorithm (MRGA)~\cite{aronson_randomized_1995} for the Maximum Matching Problem. The authors proved that the lower bound of approximation ratio of the MRGA is $\frac{1}{2}+\frac{1}{256}$ for any graph.

The network optimization problems have important roles in various applications in daily-life~\cite{demidio_clustered_2016, lin_minimum_2016},  in which some problems have clustered structure~\cite{demidio_clustered_2016}. Therefore, clustered problems are being received many interests in the scientific community. One of the newest clustered tree problem is the \gls{clustp} which is formulated by M. D'Emidio et.al.~\cite{demidio_clustered_2016}. Due to the fact that \gls{clustp} is a NP-Hard problem~\cite{demidio_clustered_2016}, the meta-heuristic algorithms are often considered to look for the optimal solutions.

Recently, two algorithms based on the \gls{mfea} are proposed to find solutions for some input graph at a time. Those algorithms encode solutions of \gls{clustp} by difference representations i.e. C-MFEA~\cite{ThanhPD_DungDA} uses the Cayley code while E-MFEA~\cite{ThanhPD_TrungTB} uses edge-set to represent the solution. The major goal of C-MFEA is to determine \gls{clustp} solution in short time while E-MFEA focuses on quality of solutions. The experimental results show that E-MFEA and C-MFEA outperform an existing algorithm. 

Although some studies have been done to produce a good solution of \gls{clustp}. However, in the aspect of finding a good solution within a small amount of time, the existing algorithms are not capable, so we have studied the combination of \glspl{rga} and \gls{spta} then applied it for the \gls{clustp}.

%% file: Sections/Sec_Proposed_Algorithm.tex
In this section, we introduce the mechanism of \gls{nrga} and explain in detail how that algorithm works through an example.

\subsection{Algorithmic skeleton of \gls{nrga}}

\begin{enumerate}
	\item \textbf{Step 1: For the root cluster}
	\begin{itemize}
		\item \textbf{Step 1.1}: Apply the Dijsktra’s algorithm from the source vertex to obtain the shortest-path tree of the root cluster.
		\item \textbf{Step 1.2}: Add the received tree in the root cluster to the solution and mark the root cluster as current cluster $V_c$.
		\item \textbf{Step 1.3}: Exclude the root cluster from the set Q.
	\end{itemize}
	\item \textbf{Step 2: For other clusters}\\
	The algorithm will add the remaining clusters one by one into the temporary solution/tree by applying the following steps:
	\begin{itemize}
		\item \textbf{Step 2.1: Finding a temporary edge connecting 2 clusters}\\ For each cluster $V_i$ left in $Q$, find an edge $(u, v)$ that connects it and the current cluster by our proposed RGA. ($u$ belongs to the current cluster, $v$ belongs to one remaining cluster $V_i \in Q$)
		\item \textbf{Step 2.2: Updating the distance between the root cluster and each cluster $V_i$:}\\ Let $w(u,v)$ be the weight of the edge connecting 2 clusters in step 1. If $dis[c] + d[root[c], u] + w[u,v] < dis[i]$, then $dis[i] = dis[c] + d[root[c], u] + w[u, v]$ and set $root[i] = v$.
		\item \textbf{Step 2.3:}
		\begin{itemize}
			\item Find a cluster $V_i \in Q$ that has the minimum value of $dis[i]$,  and then add that cluster into the temporary solution/tree using the connecting edge $(u,v)$ in step 1.
			\item The spanning tree in cluster $V_i$ is created by applying the Dijsktra’s algorithm from the local root $v$ of $V_i$ obtained in step 2.
			\item Set $V_i$ as current cluster and exclude $V_i$ from the set $Q$.
		\end{itemize}
		\item \textbf{Step 2.4:} If $Q$ is not empty, go to step 1. Otherwise, the algorithm stops. 
	\end{itemize}
\end{enumerate}

%===========----------------------Algorithmic skeleton of RGA----------------------===========
\subsection{Algorithmic skeleton of proposed RGA}
The proposed RGA is used to evaluate all clusters $V_i$ left in $Q$ for updating the distance from each cluster to the root cluster.

For each remaining cluster $V_i$ in $Q$:
\begin{itemize}
	\item Generate a random number $m$ that has the value between the number of vertices in Vi and the total number of vertices left in $Q$ $(|V_i| \leq h \leq \sum_{V_j \in Q} |V_j|$ where $V_j$ is the cluster left in $Q$, $|V_k|$ is number of vertices in cluster $k$).
	\item Let $E' = \{(u, v)\}$ be the set of all edges connecting the current cluster $V_c$ and a cluster $V_i$. Define a function:\\
	\[
		\begin{aligned}
			f \colon  E' &\to \mathbb{R}\\
			(u, v) &\mapsto h* \left(d[root[c],u]+w[u,v] \right) + costSPT(v); 
		\end{aligned}
	\]
	where $costSPT(v)$ is the total cost of the shortest paths from $v$ to other vertices in $V_i$.
	\item The probability that an edge $(u,v)$ is chosen to connect the current cluster and a cluster $V_i$ is:
	\[
		\dfrac{f(u,v)^{\gamma}}{\sum_{(u',v') \in E'} {f(u',v')}^{\gamma}}		
	\]
	where $\gamma \leq 0$ is a parameter that determines the amount of greediness.
\end{itemize}

Figure~\ref{fig:An-example-of-new-approach-to-solve-CSTP} illustrates steps of the proposed algorithm in which Figure~\ref{fig:RGA-a} presents input graph with source vertex as 1.
\begin{itemize}
	\item Figure~\ref{fig:RGA-b} illustrates the \textbf{Step 1} and \textbf{Step 2.1}, in which the sub-graph of cluster 1 is obtained after finishing \textbf{Step~1}. In \textbf{Step 2.1}, \gls{nrga} performs:
	\begin{itemize}
		\item  Consider the edges connecting the current cluster (in red) which is cluster 1 and the remaining clusters (in dashed circle).
		\item  There are 2 edges that connect cluster 1 and cluster 2 which are edge (1, 10) and edge (4, 11). Suppose \gls{nrga} chooses the edge (1, 10) as a temporary edge to connect cluster 1 and cluster 2. 
		\item  Since there is only one edge that connect cluster 1 and cluster 3, the edge (3, 5) is chosen to be a temporary edge to connect those clusters. 
	\end{itemize} 
	\item Figure~\ref{fig:RGA-c} illustrates the \textbf{Step 2.3} when:
	\begin{itemize}
		\item Since the distance between cluster 1 and cluster 2 is 5 and between cluster 1 and cluster 3 is 8 + 3 = 11, cluster 2 is set as current cluster and the edge (1, 10) is added to the temporary solution.
		\item Add the shortest-path tree of cluster 2 by the Dijkstra's algorithm with the start vertex as 10 to the temporary solution.
	\end{itemize}
	\item Figure~\ref{fig:RGA-d} illustrates the \textbf{Step 2.1} where cluster 2 is current cluster. Since the edge (13, 6) and edge (13, 15) connect cluster 2 and cluster 3, cluster 4 respectively, those edges are used to compute the distance from cluster 3 and cluster 4 to the root cluster.

	\item Figure~\ref{fig:RGA-e} illustrates the \textbf{Step 2.2} when:
	\begin{itemize}
		\item Update the distance from each unvisited cluster to root cluster (cluster 1)
		\item There are 2 edges connecting cluster 3 and the clusters that were added into the solution at previous steps (edge (3, 5) for cluster 1 and edge (6, 13) for cluster 2). Since the length of the shortest path from vertex 5 to source vertex 1 is 3 + 8 = 11 while the length of the shortest path from vertex 6 to source vertex 1 is 5 + 2 + 4 + 5 = 16, edge (3, 5) is kept to evaluate in next steps.
	\end{itemize}
	\item The process in Figure~\ref{fig:RGA-f} is similar to that of the Figure~\ref{fig:RGA-c}, in which the distance between cluster 3 and the root cluster (which is 11) is shorter than the distance between cluster 4 and the root cluster (which is 5 + 2 + 4 + 4 = 15), so cluster 3 is set as the current cluster.
	\item Figure~\ref{fig:RGA-g} show steps of \gls{nrga} which are similar to those of the previous figures. Figure~\ref{fig:RGA-g} shows the temporary solution after edge (13, 15) is selected for connecting between cluster 2 and cluster 4.
	\item Figure~\ref{fig:RGA-h} illustrates the complete solution for the \gls{clustp}.
\end{itemize}

\renewcommand{\scalefigure}{0.26}
\begin{figure*}[htbp]
	\centering
	\begin{subfigure}{.24\linewidth}
		\centering
		\includegraphics[scale=\scalefigure]{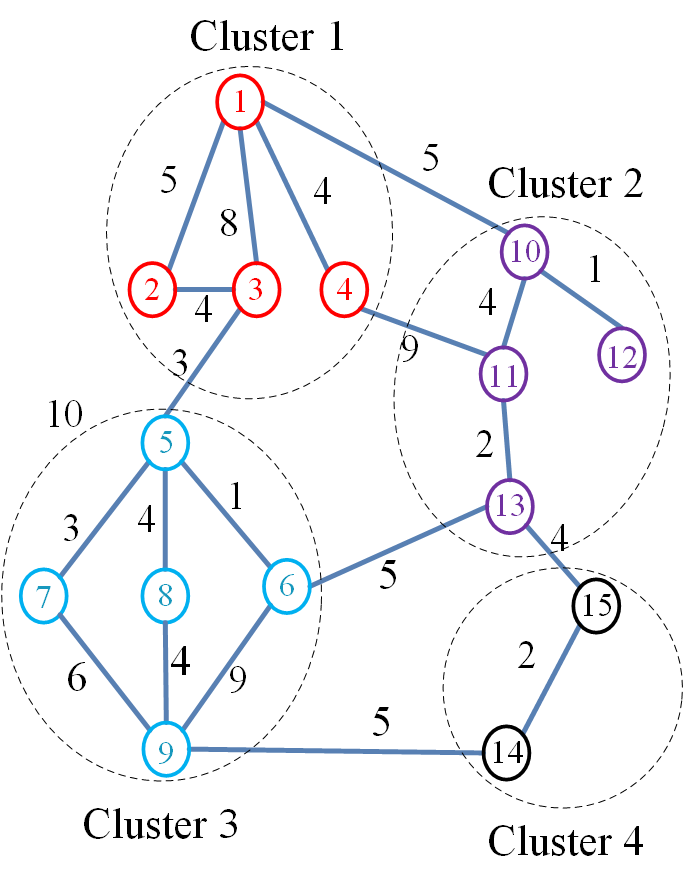}
		\caption{}
		\label{fig:RGA-a}
	\end{subfigure}
	\begin{subfigure}{.24\linewidth}
		\centering
		\includegraphics[scale=\scalefigure]{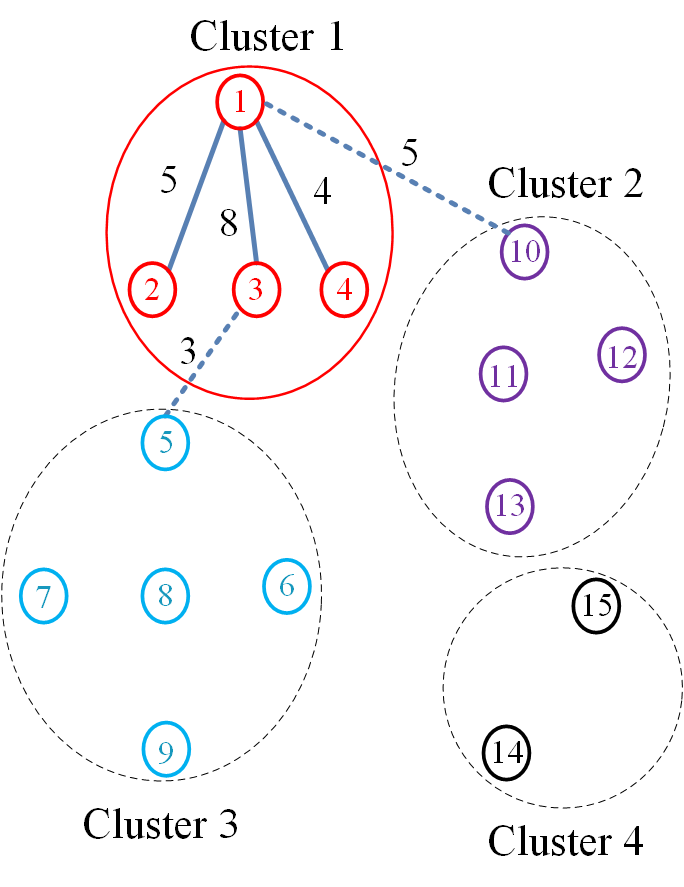}
		\caption{}
		\label{fig:RGA-b}
	\end{subfigure}
	\begin{subfigure}{.24\linewidth}
		\centering
		\includegraphics[scale=\scalefigure]{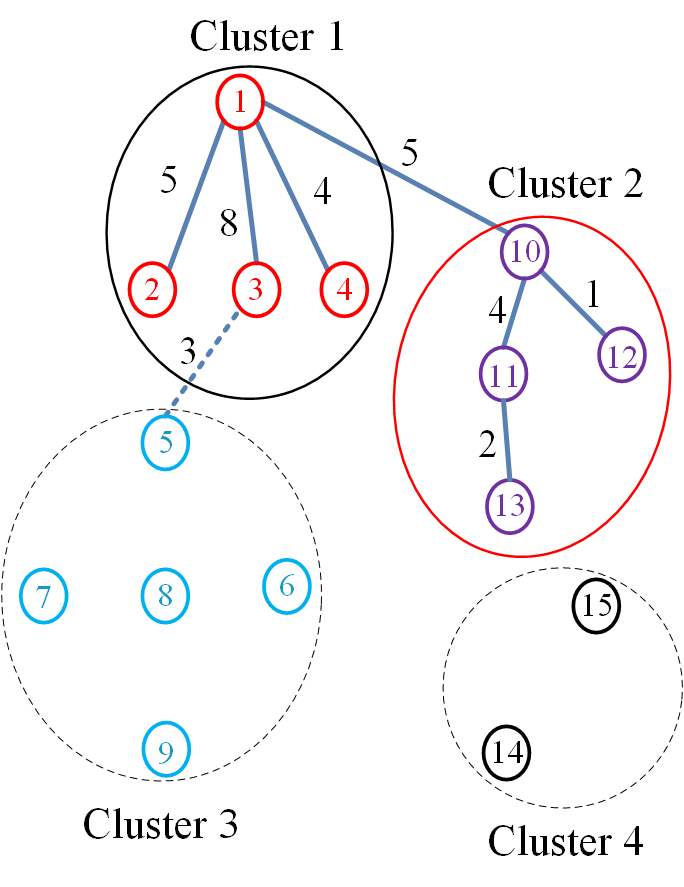}
		\caption{}
		\label{fig:RGA-c}
	\end{subfigure}
	\begin{subfigure}{.24\linewidth}
		\centering
		\includegraphics[scale=\scalefigure]{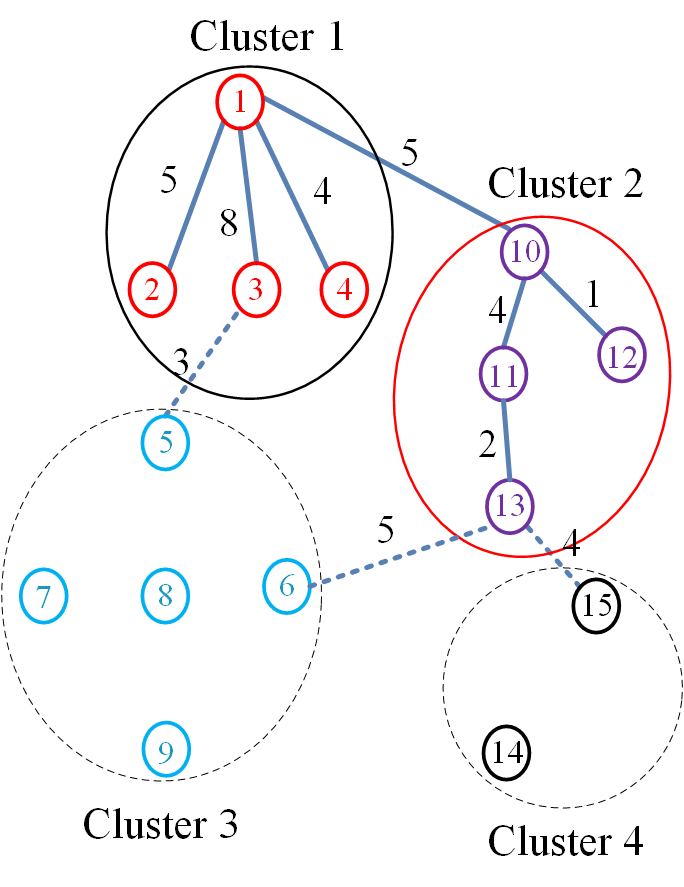}
		\caption{}
		\label{fig:RGA-d}
	\end{subfigure}
	\begin{subfigure}{.24\linewidth}
		\centering
		\includegraphics[scale=\scalefigure]{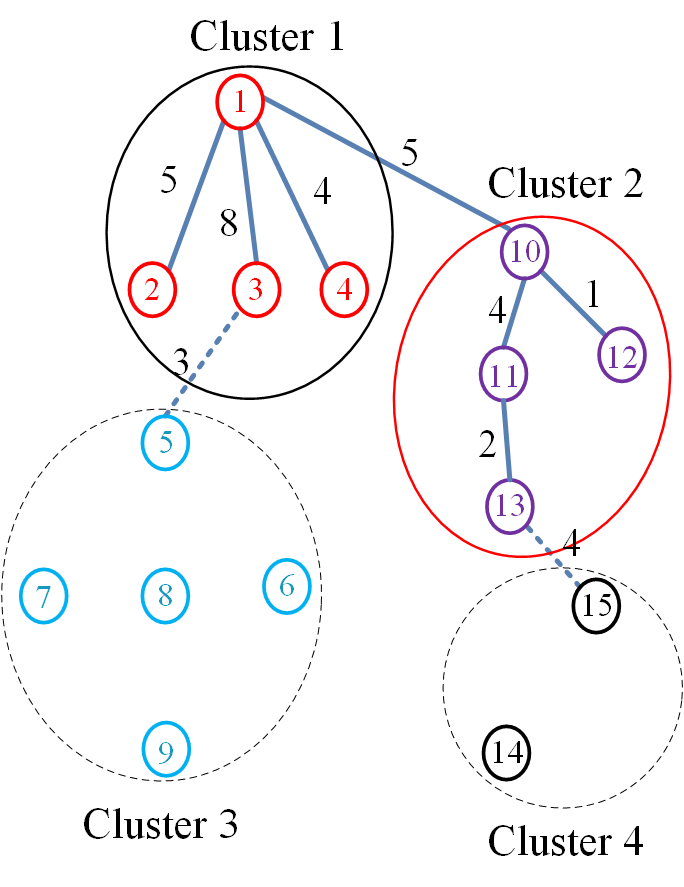}
		\caption{}
		\label{fig:RGA-e}
	\end{subfigure}
	\begin{subfigure}{.24\linewidth}
		\centering
		\includegraphics[scale=\scalefigure]{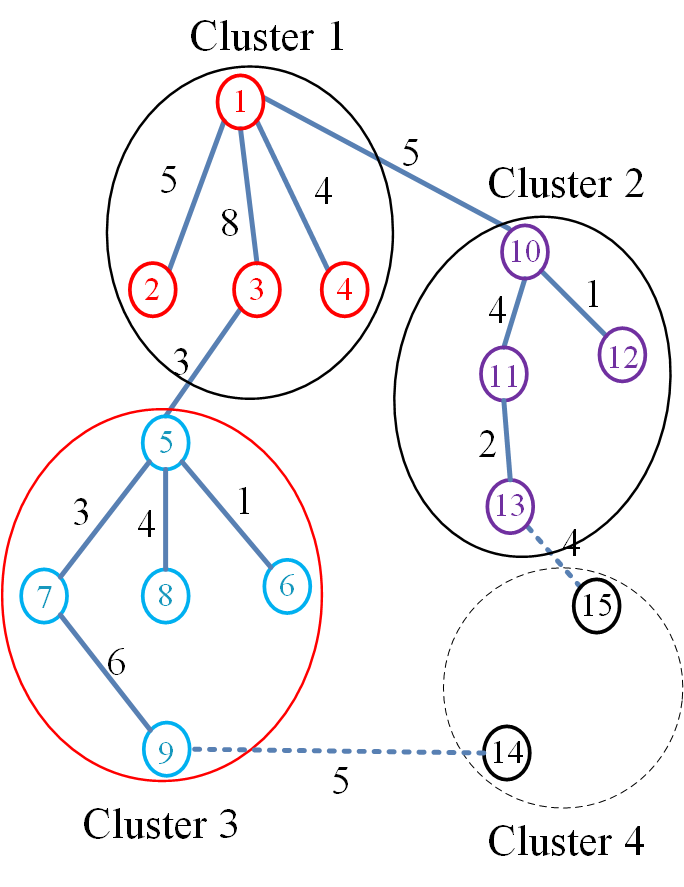}
		\caption{}
		\label{fig:RGA-f}
	\end{subfigure}
	\begin{subfigure}{.24\linewidth}
		\centering
		\includegraphics[scale=\scalefigure]{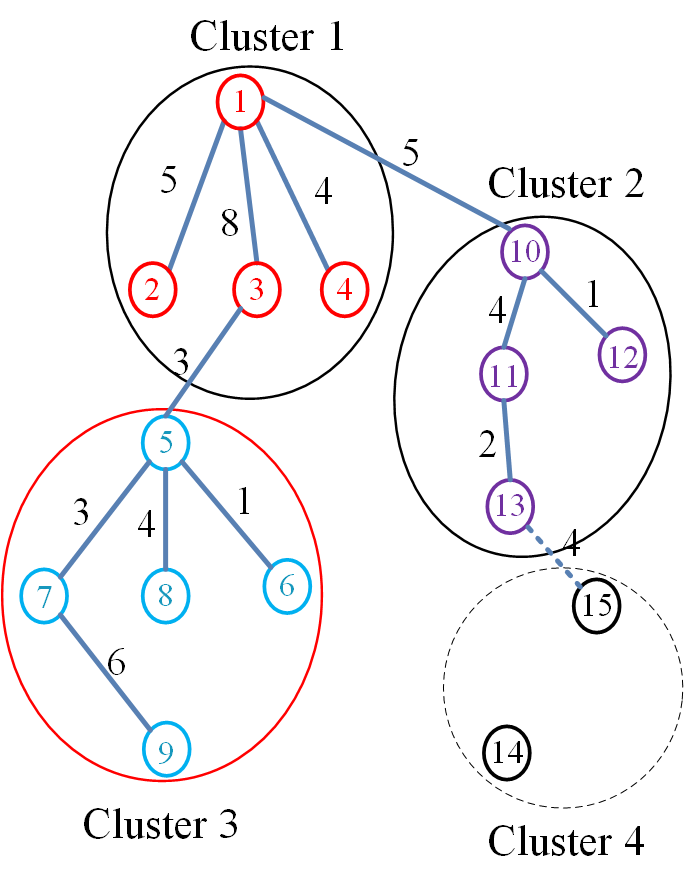}
		\caption{}
		\label{fig:RGA-g}
	\end{subfigure}
	\begin{subfigure}{.24\linewidth}
		\centering
		\includegraphics[scale=\scalefigure]{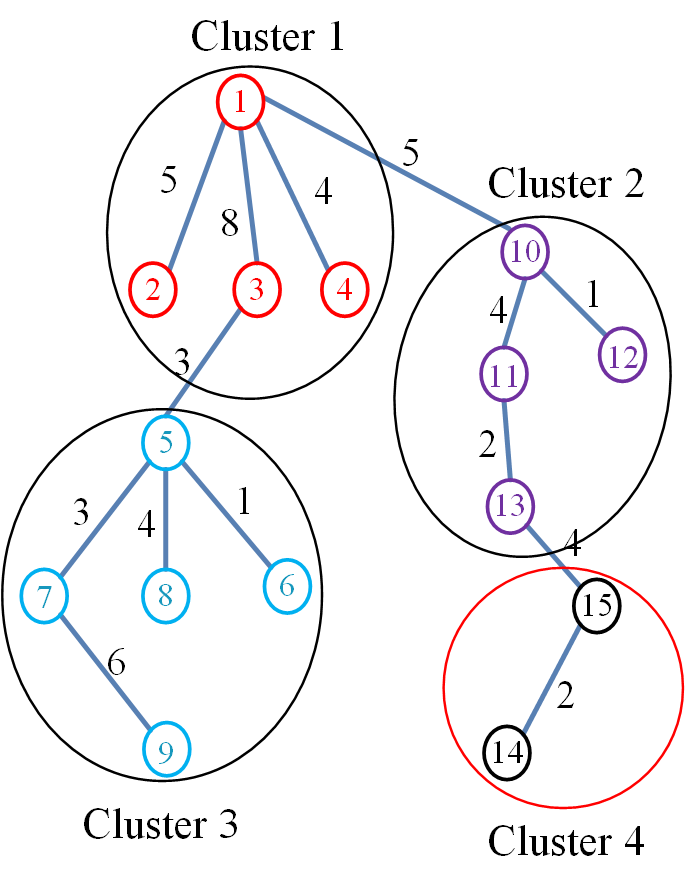}
		\caption{}
		\label{fig:RGA-h}
	\end{subfigure}
	\caption{An example of new approach to solve \gls{clustp}}
	\label{fig:An-example-of-new-approach-to-solve-CSTP}
\end{figure*}

The Pseudocode of \gls{nrga} is presented in Algorithm~\ref{alg:Structure_NRGA}.

\begin{algorithm*}
		\KwIn{Graph $G=(V,E,w)$ where $V = V_1 \cup V_2 \cup \ldots \cup V_{k}; V_p \cap V_q = \emptyset, \ \forall p \neq q$; Source vertex $s$.}
		\KwOut{A solution $T=(V_T, E_T)$}
		\BlankLine
	\Begin
	{
		$V_T \gets V$\;
		$Q \gets \{1, 2, \ldots, k\}$ \Comment{Set of clusters that have not been added to the solution}\;
		$cur\gets$ Index of cluster containing source vertex $s$ \Comment{Set current cluster as cluster containing source vertex}\;
		$T\gets Find\_Shortest\_Path\_Tree(s)$ \Comment{Construct the shortest path tree for root cluster with start vertex as s}\;
		$dis[cur]=0$\Comment{Distance from current cluster $V_{cur}$ to the root cluster}\;
		$Q\gets Q\backslash cur$\;
		\While {$Q \neq \emptyset$}
		{
			
			\ForEach {cluster $V_i$ with $i \in Q$} 
			{
				$h \gets$ random($|V_i|$, $\sum_{j \in Q} |V_j|$)\;
				\ForEach {edge $(u, v)$, $u \in V_{cur}$, $v \in V_i$} 
				{
					$d[u] \gets$ cost of the shortest path on T from the local root of $V_{cur}$ to $u$\;
					$CostSPT(v) \gets$ total cost of the shortest paths from $v$ to other vertices in $V_i$ \;
					$f(u, v) = h \times (d[u] + w[u,v]) + CostSPT(v)$\;
				}
				\BlankLine
				$Edge (a, b) \gets$ Select the fittest edge in $\left\{(u,v)|u \in V_{cur}, v \in V_i \right\}$ with probability $p(u, v) = \dfrac{f(u,v)^\gamma}{\sum_{u' \in V_{cur}, v' \in V_i} f(u', v')^\gamma}$\;
				\BlankLine
				\If {$dis[i] > dis[cur] + d[a] + w[a, b]$} 
				{
					$dis[i] = dis[cur] + d[a] + w[a,b]$\;
					$Root[i] \gets b$\;
					$temporaryEdge[i] \gets (a, b)$\;
				}
			}
			
			$cur \gets  \arg\min_{i \in Q} dis[i]$\;
			$T\gets T \cup temporaryEdge[cur]$\;
			$T\gets T \cup Find\_Shortest\_Path\_Tree(Root[cur])$\;
			$Q\gets Q \backslash cur$\;
		}
	}
	\caption{Structure of the \gls{nrga}}
	\label{alg:Structure_NRGA}
\end{algorithm*}

In Algorithm~\ref{alg:Structure_NRGA}, method $Find\_Shortest\_Path\_Tree(x)$ construct shortest path tree by using Dijkstra Algorithm with start vertex as $x$. The method $CostSPT(x)$ computes total cost of the shortest path from vertex $x$ to all vertices %which are in %same cluster %with vertex $x$.

%% file: Sections/Sec_Computational_Results.tex
\subsection{Problem instances} 
For assessment of the proposed algorithms’ performance, we created instances for the \gls{clustp} from \gls{clutsp} instances \cite{helsgaun_solving_2011}\cite{mestria_grasp_2013} by adding the information of the source vertex. The main reason for building \gls{clustp} instances from \gls{clutsp}  instances was that \gls{clutsp}  instances had been generated with various algorithms to be suitable for clustered problems \cite{mestria_grasp_2013}. 

For evaluation of the proposed algorithms, small instances of \gls{clutsp} were selected for the building of instances for \gls{clustp}.

All tested instances are available via~\cite{Pham_Dinh_Thanh_2018_Instances}.
\subsection{Experimental setup}
We focused on the following criteria to assess the quality of the output of the algorithms.
\begin{center}
	\begin{tabular}{p{2.5cm}p{5.5cm}}
		\hline 
		\multicolumn{2}{c}{Criteria} \\ 
		\hline 
		Average (Avg)	 & Average function value over all runs \\ 
		\hline 
		Best-found (BF) & Best function value achieved over all runs \\ 
		\hline 
	\end{tabular} 
\end{center}

To compare the effectiveness of the two algorithms, we evaluated the difference between the costs of the results obtained by the two algorithms A and B by the following formula:

\begin{center}
	$PI(A,B)=\dfrac{C_B-C_A}{C_B}*100\%$
\end{center}
where $C_A$ and $C_B$ denote the costs of the best solutions generated by A and B respectively.

To evaluate the performance of the \gls{mfea} in solving the \gls{clustp}, we implemented two sets of experiments.
\begin{itemize}
	\item[$\bullet$] In the first set, the solution's qualities by the \text{C-MFEA}~\cite{ThanhPD_DungDA} and E-MFEA~\cite{ThanhPD_TrungTB} on each instance were compared with the corresponding values by the \gls{nrga}.	
	\item[$\bullet$] In the second set, various experiments were performed to analyze possible influencing factors.    
\end{itemize}

Each problem instance was evaluated 30 times on Intel Core i7-3.60GHz, 16GB RAM computer using Windows 8 64-bit. The source codes were written in the Visual C\# language.

%The simulation parameters include population size = 100, number of evaluations = 50000, probability of random mating~=~0.5, mutation rate = 0.05 and number of tasks = 2.

\subsection{Experimental results}
\subsubsection{Analysis of the parameter $\gamma$}
Table~\ref{tab:GammaType1} demonstrates the results of \gls{nrga} with regards to different values of $\gamma$. These results indicate that, for complete graph, larger values of $\gamma$ tend to give better results. In other words, the more greedy our algorithm becomes, the better results it would produce.

Hence, in this paper, we select the results when $\gamma$ = 50 to compare our heuristic with some other existing algorithms.

\subsubsection{Comparison between the performance of existing algorithms and that of the proposed algorithm}
In this section, the comparison between the results received by the \gls{nrga} and those of 2 other existing algorithms (\text{E-MFEA} and C-MFEA) will be discussed. Table~\ref{tab:ResultsType1},~~\ref{tab:ResultsType5}~and~~\ref{tab:ResultsType6} illustrate the results achieved by all 3 algorithms on instances of Type 1, Type 5 and Type 6 respectively. In those tables, the symbol "-" indicates that the corresponding instances are not considered.

In general, the results obtained by \gls{nrga} were considerably better in quality than those obtained by the C-MFEA and the E-MFEA in all test cases of the Type 1, Type 5 and Type 6.

As is observed in Table~\ref{tab:ResultsType1}, the proposed algorithm exceeded the E-MFEA on all instances in Type 1 with an average PI(\gls{nrga},~E-MFEA) of 15.9\%. In particluar, the highest PI(\gls{nrga},~E-MFEA) was 39.2\% on instance 10st70.  Results obtained by the 2 algorithms on instances in Type 5 and Type 6 are presented in Table~\ref{tab:ResultsType5},~\ref{tab:ResultsType6}. The average PI(\gls{nrga},~\text{E-MFEA}) was under one fifth on both types with the highest PI(\gls{nrga},~E-MFEA) of Type 5 and Type 6 were 32.8\% and 37.9\% respectively.

In comparison with the C-MFEA, it is clear that the \gls{nrga} outperformed it significantly on all test cases. The average PI(\gls{nrga},~C-MFEA) obtained from each of the 3 types was above 40\% with the greatest PI(\gls{nrga},~C-MFEA) was 64.5\% on instance 4pr76-2x2 in Type 6.

%% file: Sections/Sec_Conclusion.tex
In this paper, we proposed a new heuristic based on the combination between \gls{rga} and \gls{spta} to solve the \gls{clustp}. To evaluate its effectiveness, the heuristic has been experimented on multiple types of Euclidean instances. The results have strongly proven the efficiency of our proposed heuristic.
In the future, we will continue to analyze the results of the heuristic for many other types of graphs (Non-Euclidean, Sparse graph). Furthermore, we will add the shortest path tree cost factor into the formula evaluating the path from the root cluster to another cluster, in order to optimize the the cost of its spanning tree.

%% file: Table_Data/GammaType1.tex
% Table generated by Excel2LaTeX from sheet 'SoSanhCEC2018 (3)'
\begin{table*}[htbp]
  \centering
  \caption{Results Obtained by \gls{nrga} with different parameter setting on Instances In Type 1}
    \begin{tabular}{|l|r|r|r|r|r|r|r|r|r|r|r|r|r|r|}
    \hline
    \multicolumn{1}{|c|}{\multirow{2}[4]{*}{\textbf{Instances}}} & \multicolumn{2}{c|}{$\gamma$ = 1} & \multicolumn{2}{c|}{$\gamma$ = 5} & \multicolumn{2}{c|}{$\gamma$ = 10} & \multicolumn{2}{c|}{$\gamma$ = 20} & \multicolumn{2}{c|}{$\gamma$ = 30} & \multicolumn{2}{c|}{$\gamma$ = 40} & \multicolumn{2}{c|}{$\gamma$ = 50} \\
\cline{2-15}          & \multicolumn{1}{c|}{\textbf{BF}} & \multicolumn{1}{c|}{\textbf{Avg}} & \multicolumn{1}{c|}{\textbf{BF}} & \multicolumn{1}{c|}{\textbf{Avg}} & \multicolumn{1}{c|}{\textbf{BF}} & \multicolumn{1}{c|}{\textbf{Avg}} & \multicolumn{1}{c|}{\textbf{BF}} & \multicolumn{1}{c|}{\textbf{Avg}} & \multicolumn{1}{c|}{\textbf{BF}} & \multicolumn{1}{c|}{\textbf{Avg}} & \multicolumn{1}{c|}{\textbf{BF}} & \multicolumn{1}{c|}{\textbf{Avg}} & \multicolumn{1}{c|}{\textbf{BF}} & \multicolumn{1}{c|}{\textbf{Avg}} \\
    \hline
    10berlin52 & 45828.6 & 48866.6 & 44303.5 & 46428.2 & 44042.3 & 45325.5 & 43802.7 & 44500.4 & 43724.1 & 44214.5 & 43738.6 & 44002.0 & 43738.6 & 43971.0 \\
    \hline
    10eil51 & 2009.8 & 2297.3 & 1733.9 & 1837.8 & 1713.3 & 1760.9 & 1713.2 & 1729.4 & 1713.2 & 1725.1 & 1713.2 & 1723.8 & 1713.2 & 1723.2 \\
    \hline
    10eil76 & 2596.4 & 3562.5 & 2249.4 & 2426.2 & 2204.0 & 2254.1 & 2203.3 & 2217.7 & 2203.3 & 2210.1 & 2203.3 & 2209.1 & 2203.3 & 2208.4 \\
    \hline
    10kroB100 & 160439.5 & 188533.8 & 148188.4 & 161012.6 & 145243.6 & 150565.0 & 141824.9 & 145118.9 & 140820.0 & 142994.3 & 140653.5 & 142257.6 & 140635.1 & 141951.4 \\
    \hline
    10pr76 & 605003.0 & 674451.2 & 565285.7 & 592198.8 & 532573.4 & 555253.6 & 523893.3 & 534672.9 & 523249.0 & 529046.4 & 522572.2 & 527656.6 & 522572.2 & 525733.1 \\
    \hline
    10rat99 & 9008.3 & 10210.9 & 7925.8 & 8466.1 & 7621.9 & 7888.7 & 7520.2 & 7653.7 & 7520.2 & 7603.2 & 7520.2 & 7576.8 & 7520.2 & 7562.1 \\
    \hline
    10st70 & 3543.5 & 4335.9 & 3174.3 & 3467.3 & 3117.0 & 3219.3 & 3099.2 & 3151.2 & 3099.5 & 3139.2 & 3099.5 & 3132.6 & 3099.5 & 3131.7 \\
    \hline
    15berlin52 & 28984.6 & 32028.9 & 26974.8 & 28433.5 & 26522.0 & 27214.6 & 26346.0 & 26662.9 & 26345.5 & 26543.2 & 26312.1 & 26458.3 & 26312.0 & 26437.7 \\
    \hline
    15eil51 & 1492.4 & 1632.2 & 1355.7 & 1436.7 & 1312.3 & 1359.2 & 1309.9 & 1325.6 & 1306.8 & 1318.5 & 1306.7 & 1317.3 & 1306.7 & 1313.8 \\
    \hline
    15eil76 & 3236.2 & 3503.8 & 2993.3 & 3137.4 & 2957.2 & 3011.0 & 2918.3 & 2955.9 & 2914.4 & 2933.6 & 2911.3 & 2925.3 & 2911.3 & 2921.8 \\
    \hline
    15pr76 & 771109.3 & 816189.0 & 732956.0 & 755924.1 & 718439.4 & 730875.2 & 708266.2 & 716625.6 & 707937.9 & 712899.3 & 706060.3 & 710154.1 & 705017.3 & 708944.9 \\
    \hline
    15st70 & 4504.0 & 4826.7 & 4294.8 & 4452.4 & 4208.0 & 4311.8 & 4152.5 & 4207.8 & 4134.7 & 4177.1 & 4130.9 & 4155.6 & 4129.9 & 4147.3 \\
    \hline
    25eil101 & 4901.3 & 5081.4 & 4715.2 & 4775.5 & 4698.0 & 4721.7 & 4683.9 & 4694.1 & 4683.4 & 4688.9 & 4680.8 & 4687.5 & 4680.8 & 4686.1 \\
    \hline
    25kroA100 & 161039.3 & 172263.7 & 151282.7 & 157622.0 & 149080.1 & 151413.1 & 147698.6 & 148699.5 & 147430.1 & 148123.1 & 147300.8 & 147901.0 & 147239.0 & 147716.8 \\
    \hline
    25lin105 & 107095.9 & 114122.6 & 102464.3 & 105958.3 & 99662.0 & 101742.0 & 98523.0 & 99646.7 & 98372.8 & 98946.4 & 98222.0 & 98680.4 & 98087.5 & 98502.9 \\
    \hline
    25rat99 & 7558.1 & 8076.6 & 7119.2 & 7256.1 & 6924.0 & 7035.6 & 6884.7 & 6927.9 & 6846.3 & 6888.3 & 6846.3 & 6875.6 & 6846.3 & 6867.8 \\
    \hline
    50eil101 & 3918.5 & 4012.2 & 3839.5 & 3876.7 & 3830.1 & 3841.1 & 3828.1 & 3831.6 & 3827.3 & 3829.5 & 3827.3 & 3828.8 & 3827.3 & 3828.1 \\
    \hline
    50kroA100 & 164224.3 & 167998.7 & 161519.8 & 163461.6 & 160527.9 & 161367.2 & 159947.6 & 160412.2 & 159835.0 & 160160.7 & 159815.2 & 160069.3 & 159815.2 & 160029.9 \\
    \hline
    50kroB100 & 136086.6 & 138003.5 & 134193.1 & 135198.2 & 133587.6 & 134059.0 & 133264.4 & 133554.4 & 133247.7 & 133419.1 & 133181.9 & 133362.9 & 133135.4 & 133325.8 \\
    \hline
    50lin105 & 146874.9 & 148362.1 & 146203.2 & 147101.1 & 146131.9 & 146531.3 & 145952.5 & 146187.5 & 145929.5 & 146046.2 & 145904.2 & 145985.5 & 145869.9 & 145951.8 \\
    \hline
    50rat99 & 8238.0 & 8506.7 & 8056.1 & 8143.1 & 8031.7 & 8059.4 & 8011.6 & 8026.2 & 8008.7 & 8020.1 & 8010.7 & 8017.9 & 8010.6 & 8016.8 \\
    \hline
    5berlin52 & 27294.5 & 34028.5 & 24331.1 & 27211.5 & 23114.1 & 24797.2 & 22746.4 & 23745.3 & 22746.4 & 23256.4 & 22746.4 & 23124.6 & 22746.4 & 23106.9 \\
    \hline
    5eil51 & 2122.9 & 2753.7 & 1841.2 & 2039.1 & 1784.2 & 1871.2 & 1776.6 & 1826.8 & 1773.3 & 1804.0 & 1770.5 & 1792.0 & 1770.5 & 1792.3 \\
    \hline
    5eil76 & 3048.4 & 3978.7 & 2711.5 & 3182.5 & 2637.3 & 2849.6 & 2630.9 & 2699.6 & 2630.9 & 2690.6 & 2630.9 & 2672.4 & 2630.8 & 2658.4 \\
    \hline
    5pr76 & 769591.9 & 1012706.6 & 603094.7 & 686969.3 & 594661.7 & 631083.5 & 585155.6 & 605210.0 & 585008.0 & 595592.2 & 585008.0 & 591396.6 & 585008.0 & 589778.1 \\
    \hline
    5st70 & 5315.9 & 6077.4 & 4584.4 & 5181.5 & 4596.0 & 4902.2 & 4542.6 & 4660.5 & 4530.2 & 4598.0 & 4525.1 & 4572.0 & 4520.1 & 4562.8 \\
    \hline
    \end{tabular}%
  \label{tab:GammaType1}%
\end{table*}%

%% file: Table_Data/ResultsType1.tex
% Table generated by Excel2LaTeX from sheet 'Compare with CEC 2018_v2'
\begin{table*}[htbp]
  \centering
  \caption{Results Obtained by E-MFEA, C-MFEA and \gls{nrga} on Instances In Type 1}
    \begin{tabular}{|l|r|r|c|r|r|c|l|l|l|}
    \hline
    \multicolumn{1}{|c|}{\multirow{2}[4]{*}{\textbf{Instances}}} & \multicolumn{3}{c|}{\textbf{E-MFEA}} & \multicolumn{3}{c|}{\textbf{C-MFEA}} & \multicolumn{3}{c|}{\textbf{\gls{nrga}}} \\
\cline{2-10}          & \multicolumn{1}{c|}{\textbf{BF}} & \multicolumn{1}{c|}{\textbf{Avg}} & \multicolumn{1}{p{4.445em}|}{\textbf{Time (s)}} & \multicolumn{1}{c|}{\textbf{BF}} & \multicolumn{1}{c|}{\textbf{Avg}} & \multicolumn{1}{p{3.555em}|}{\textbf{Time (s)}} & \multicolumn{1}{c|}{\textbf{BF}} & \multicolumn{1}{c|}{\textbf{Avg}} & \multicolumn{1}{p{4.165em}|}{\textbf{Time (s)}} \\
    \hline
    10berlin52 & 45684.5 & 47075.7 & 0.10  & -     & -     & -     & \multicolumn{1}{r|}{43738.6} & \multicolumn{1}{r|}{43971.0} & \multicolumn{1}{r|}{0.00} \\
    \hline
    10eil51 & 1923.9 & 2020.3 & 0.10  & 3027.7 & 3513.3 & 0.05  & \multicolumn{1}{r|}{1713.2} & \multicolumn{1}{r|}{1723.2} & \multicolumn{1}{r|}{0.00} \\
    \hline
    10eil76 & 3089.3 & 3418.7 & 0.22  & 4263.1 & 5175.2 & 0.05  & \multicolumn{1}{r|}{2203.3} & \multicolumn{1}{r|}{2208.4} & \multicolumn{1}{r|}{0.00} \\
    \hline
    10kroB100 & 203481.4 & 221058.2 & 0.22  & 301982.8 & 363749.0 & 0.08  & \multicolumn{1}{r|}{140635.1} & \multicolumn{1}{r|}{141951.4} & \multicolumn{1}{r|}{0.00} \\
    \hline
    10pr76 & 656800.9 & 685310.3 & 0.22  & 914315.2 & 1159871.7 & 0.08  & \multicolumn{1}{r|}{522572.2} & \multicolumn{1}{r|}{525733.1} & \multicolumn{1}{r|}{0.00} \\
    \hline
    10rat99 & 10381.7 & 11015.9 & 0.22  & 13954.6 & 17406.1 & 0.07  & \multicolumn{1}{r|}{7520.2} & \multicolumn{1}{r|}{7562.1} & \multicolumn{1}{r|}{0.00} \\
    \hline
    10st70 & 5723.3 & 5833.4 & 0.17  & -     & -     & -     & \multicolumn{1}{r|}{3099.5} & \multicolumn{1}{r|}{3131.7} & \multicolumn{1}{r|}{0.00} \\
    \hline
    15berlin52 & 29246.2 & 30279.9 & 0.17  & -     & -     & -     & \multicolumn{1}{r|}{26312.0} & \multicolumn{1}{r|}{26437.7} & \multicolumn{1}{r|}{0.00} \\
    \hline
    15eil51 & 1753.5 & 1954.1 & 0.15  & -     & -     & -     & \multicolumn{1}{r|}{1306.7} & \multicolumn{1}{r|}{1313.8} & \multicolumn{1}{r|}{0.00} \\
    \hline
    15eil76 & 3374.9 & 3452.4 & 0.15  & -     & -     & -     & \multicolumn{1}{r|}{2911.3} & \multicolumn{1}{r|}{2921.8} & \multicolumn{1}{r|}{0.00} \\
    \hline
    15pr76 & 772012.8 & 796271.2 & 0.15  & -     & -     & -     & \multicolumn{1}{r|}{705017.3} & \multicolumn{1}{r|}{708944.9} & \multicolumn{1}{r|}{0.00} \\
    \hline
    15st70 & 4921.2 & 5308.3 & 0.15  & -     & -     & -     & \multicolumn{1}{r|}{4129.9} & \multicolumn{1}{r|}{4147.3} & \multicolumn{1}{r|}{0.00} \\
    \hline
    25eil101 & 5241.4 & 5384.7 & 0.27  & -     & -     & -     & \multicolumn{1}{r|}{4680.8} & \multicolumn{1}{r|}{4686.1} & \multicolumn{1}{r|}{0.00} \\
    \hline
    25kroA100 & 165880.9 & 169702.2 & 0.27  & -     & -     & -     & \multicolumn{1}{r|}{147239.0} & \multicolumn{1}{r|}{147716.8} & \multicolumn{1}{r|}{0.00} \\
    \hline
    25lin105 & 107677.9 & 110598.5 & 0.30  & -     & -     & -     & \multicolumn{1}{r|}{98087.5} & \multicolumn{1}{r|}{98502.9} & \multicolumn{1}{r|}{0.00} \\
    \hline
    25rat99 & 9464.2 & 9690.5 & 0.30  & -     & -     & -     & \multicolumn{1}{r|}{6846.3} & \multicolumn{1}{r|}{6867.8} & \multicolumn{1}{r|}{0.00} \\
    \hline
    50eil101 & 4239.2 & 4459.4 & 0.37  & -     & -     & -     & \multicolumn{1}{r|}{3827.3} & \multicolumn{1}{r|}{3828.1} & \multicolumn{1}{r|}{0.00} \\
    \hline
    50kroA100 & 180990.7 & 199637.3 & 0.37  & -     & -     & -     & \multicolumn{1}{r|}{159815.2} & \multicolumn{1}{r|}{160029.9} & \multicolumn{1}{r|}{0.00} \\
    \hline
    50kroB100 & 156209.3 & 170468.1 & 0.37  & -     & -     & -     & \multicolumn{1}{r|}{133135.4} & \multicolumn{1}{r|}{133325.8} & \multicolumn{1}{r|}{0.00} \\
    \hline
    50lin105 & 153465.7 & 158775.5 & 0.37  & -     & -     & -     & \multicolumn{1}{r|}{145869.9} & \multicolumn{1}{r|}{145951.8} & \multicolumn{1}{r|}{0.00} \\
    \hline
    50rat99 & 9747.3 & 11328.3 & 0.82  & -     & -     & -     & \multicolumn{1}{r|}{8010.6} & \multicolumn{1}{r|}{8016.8} & \multicolumn{1}{r|}{0.00} \\
    \hline
    5berlin52 & 35387.5 & 37595.9 & 0.82  & 42296.7 & 48591.5 & 0.07  & \multicolumn{1}{r|}{22746.4} & \multicolumn{1}{r|}{23106.9} & \multicolumn{1}{r|}{0.00} \\
    \hline
    5eil51 & 2101.3 & 2367.0 & 0.18  & 2380.1 & 2691.7 & 0.05  & \multicolumn{1}{r|}{1770.5} & \multicolumn{1}{r|}{1792.3} & \multicolumn{1}{r|}{0.00} \\
    \hline
    5eil76 & 3450.1 & 3688.0 & 0.18  & 4962.0 & 5583.6 & 0.05  & \multicolumn{1}{r|}{2630.8} & \multicolumn{1}{r|}{2658.4} & \multicolumn{1}{r|}{0.00} \\
    \hline
    5pr76 & 709511.2 & 799642.4 & 0.23  & 1056191.9 & 1261431.3 & 0.05  & \multicolumn{1}{r|}{585008.0} & \multicolumn{1}{r|}{589778.1} & \multicolumn{1}{r|}{0.00} \\
    \hline
    5st70 & 5430.2 & 5693.8 & 0.23  & 6598.6 & 7550.2 & 0.05  & \multicolumn{1}{r|}{4520.1} & \multicolumn{1}{r|}{4562.8} & \multicolumn{1}{r|}{0.00} \\
    \hline
    \end{tabular}%
  \label{tab:ResultsType1}
\end{table*}%

%% file: Table_Data/ResultsType5.tex
% Table generated by Excel2LaTeX from sheet 'Compare with CEC 2018_v2'
\begin{table}[htbp]
  \centering
  \caption{Results Obtained by E-MFEA, C-MFEA and \gls{nrga} on Instances In Type 5}
    \begin{tabular}{|l|r|r|c|r|r|c|r|r|r|}
    \hline
    \multicolumn{1}{|c|}{\multirow{2}[4]{*}{\textbf{Instances}}} & \multicolumn{3}{c|}{\textbf{E-MFEA}} & \multicolumn{3}{c|}{\textbf{C-MFEA}} & \multicolumn{3}{c|}{\textbf{\gls{nrga}}} \\
\cline{2-10}          & \multicolumn{1}{c|}{\textbf{BF}} & \multicolumn{1}{c|}{\textbf{Avg}} & \multicolumn{1}{p{4.445em}|}{\textbf{Time (s)}} & \multicolumn{1}{c|}{\textbf{BF}} & \multicolumn{1}{c|}{\textbf{Avg}} & \multicolumn{1}{p{3.555em}|}{\textbf{Time (s)}} & \multicolumn{1}{c|}{\textbf{BF}} & \multicolumn{1}{c|}{\textbf{Avg}} & \multicolumn{1}{p{4.165em}|}{\textbf{Time (s)}} \\
    \hline
    10i120-46 & 122349.2 & 125510.1 & 0.32  & 156097.1 & 184275.7 & 0.10  & 94055.2 & 94596.7 & 0.00 \\
    \hline
    10i30-17 & 14718.5 & 15740.9 & 0.32  & -     & -     & -     & 13276.6 & 13289.2 & 0.00 \\
    \hline
    10i45-18 & 27068.0 & 29306.5 & 0.12  & 37121.1 & 42932.6 & 0.10  & 23267.6 & 23344.8 & 0.00 \\
    \hline
    10i60-21 & 42386.6 & 44667.3 & 0.12  & 53877.1 & 68825.7 & 0.05  & 33744.5 & 35002.0 & 0.00 \\
    \hline
    10i65-21 & 47165.1 & 50815.3 & 0.15  & 66268.0 & 79374.5 & 0.05  & 37386.7 & 37677.0 & 0.00 \\
    \hline
    10i70-21 & 48058.1 & 51889.0 & 0.15  & 61048.5 & 76907.1 & 0.08  & 38543.8 & 38855.0 & 0.00 \\
    \hline
    10i75-22 & 74952.5 & 77605.4 & 0.25  & -     & -     & -     & 65411.9 & 65783.1 & 0.00 \\
    \hline
    10i90-33 & 66438.4 & 67881.2 & 0.25  & 81379.3 & 97534.6 & 0.08  & 52091.2 & 52617.6 & 0.00 \\
    \hline
    5i120-46 & 92826.2 & 103713.6 & 0.47  & -     & -     & -     & 61776.0 & 62393.2 & 0.00 \\
    \hline
    5i30-17 & 15801.5 & 17664.7 & 0.47  & -     & -     & -     & 14399.9 & 14399.9 & 0.00 \\
    \hline
    5i45-18 & 19813.0 & 23639.3 & 0.12  & 24131.3 & 27649.3 & 0.05  & 14884.3 & 14893.0 & 0.00 \\
    \hline
    5i60-21 & 36445.8 & 39060.5 & 0.12  & 48867.1 & 57579.8 & 0.05  & 28422.7 & 28584.2 & 0.00 \\
    \hline
    5i65-21 & 38682.7 & 41488.6 & 0.15  & 51818.2 & 62189.7 & 0.07  & 31244.3 & 31684.7 & 0.00 \\
    \hline
    5i70-21 & 50025.1 & 54839.7 & 0.15  & 69390.4 & 82771.9 & 0.07  & 35052.8 & 35384.4 & 0.00 \\
    \hline
    5i75-22 & 41260.0 & 49758.6 & 0.27  & 66131.0 & 78693.8 & 0.08  & 34811.1 & 34993.8 & 0.00 \\
    \hline
    5i90-33 & 69640.4 & 74725.8 & 0.27  & 91746.3 & 99266.9 & 0.08  & 52128.9 & 52916.0 & 0.00 \\
    \hline
    7i30-17 & 24546.7 & 26344.9 & 0.10  & -     & -     & -     & 20438.9 & 20450.0 & 0.00 \\
    \hline
    7i45-18 & 32673.8 & 34086.1 & 0.10  & -     & -     & -     & 20512.0 & 20973.8 & 0.00 \\
    \hline
    7i60-21 & 45073.1 & 48395.9 & 0.15  & 55312.7 & 67556.5 & 0.05  & 36263.9 & 36339.5 & 0.00 \\
    \hline
    7i65-21 & 47276.4 & 49872.8 & 0.15  & 58179.2 & 71715.7 & 0.05  & 34847.6 & 34881.9 & 0.00 \\
    \hline
    7i70-21 & 54019.4 & 60450.6 & 13.47 & 57464.9 & 67043.4 & 1.10  & 39757.8 & 39970.3 & 0.00 \\
    \hline
    \end{tabular}%
  	\label{tab:ResultsType5}%
\end{table}%

%% file: Table_Data/ResultsType6.tex
% Table generated by Excel2LaTeX from sheet 'Compare with CEC 2018_v2'
\begin{table*}[htbp]
  \centering
  \caption{Results Obtained by E-MFEA, C-MFEA and \gls{nrga} on Instances In Type 6}
    \begin{tabular}{|l|r|r|c|r|r|c|r|r|r|}
    \hline
    \multicolumn{1}{|c|}{\multirow{2}[4]{*}{\textbf{Instances}}} & \multicolumn{3}{c|}{\textbf{E-MFEA}} & \multicolumn{3}{c|}{\textbf{C-MFEA}} & \multicolumn{3}{c|}{\textbf{\gls{nrga}}} \\
\cline{2-10}          & \multicolumn{1}{c|}{\textbf{BF}} & \multicolumn{1}{c|}{\textbf{Avg}} & \multicolumn{1}{p{4.445em}|}{\textbf{Time (s)}} & \multicolumn{1}{c|}{\textbf{BF}} & \multicolumn{1}{c|}{\textbf{Avg}} & \multicolumn{1}{p{3.555em}|}{\textbf{Time (s)}} & \multicolumn{1}{c|}{\textbf{BF}} & \multicolumn{1}{c|}{\textbf{Avg}} & \multicolumn{1}{p{4.165em}|}{\textbf{Time (s)}} \\
    \hline
    10berlin52-2x5 & 34749.2 & 36828.8 & 1.60  & -     & -     & -     & 27472.4 & 27723.2 & 0.00 \\
    \hline
    12eil51-3x4 & 1922.4 & 2000.3 & 0.18  & 3115.8 & 3648.4 & 0.07  & 1699.1 & 1702.3 & 0.00 \\
    \hline
    12eil76-3x4 & 3197.7 & 3330.0 & 0.18  & 5219.5 & 6381.5 & 0.07  & 2650.8 & 2653.2 & 0.00 \\
    \hline
    12pr76-3x4 & 699229.9 & 723373.3 & 0.22  & -     & -     & -     & 600597.6 & 603474.4 & 0.00 \\
    \hline
    12st70-3x4 & 5113.1 & 5431.3 & 0.22  & 6976.2 & 8579.5 & 0.12  & 4128.1 & 4144.6 & 0.00 \\
    \hline
    15pr76-3x5 & 560767.5 & 576792.5 & 0.23  & -     & -     & -     & 526596.7 & 532896.8 & 0.00 \\
    \hline
    16eil51-4x4 & 1459.1 & 1490.5 & 0.23  & -     & -     & -     & 1302.4 & 1304.0 & 0.00 \\
    \hline
    16eil76-4x4 & 3321.5 & 3453.1 & 0.38  & -     & -     & -     & 2042.4 & 2053.8 & 0.00 \\
    \hline
    16lin105-4x4 & 158387.1 & 162660.9 & 0.38  & -     & -     & -     & 125052.2 & 125685.3 & 0.00 \\
    \hline
    16st70-4x4 & 3410.2 & 3519.6 & 0.25  & -     & -     & -     & 2939.3 & 2966.1 & 0.00 \\
    \hline
    18pr76-3x6 & 735572.7 & 771984.5 & 0.25  & -     & -     & -     & 642733.0 & 646399.1 & 0.00 \\
    \hline
    20eil51-4x5 & 2583.1 & 2621.0 & 0.22  & -     & -     & -     & 2284.2 & 2287.9 & 0.00 \\
    \hline
    20eil76-4x5 & 2886.0 & 3038.8 & 0.22  & -     & -     & -     & 2385.9 & 2392.6 & 0.00 \\
    \hline
    20st70-4x5 & 4387.9 & 4582.7 & 0.38  & -     & -     & -     & 2939.4 & 2945.7 & 0.00 \\
    \hline
    25eil101-5x5 & 4314.2 & 4546.4 & 0.38  & -     & -     & -     & 3609.1 & 3622.0 & 0.00 \\
    \hline
    25eil51-5x5 & 1550.4 & 1640.2 & 0.22  & -     & -     & -     & 1474.6 & 1476.2 & 0.00 \\
    \hline
    25eil76-5x5 & 3523.0 & 3722.1 & 0.22  & -     & -     & -     & 2193.1 & 2194.8 & 0.00 \\
    \hline
    25rat99-5x5 & 12283.6 & 12635.4 & 0.45  & -     & -     & -     & 11400.3 & 11418.9 & 0.00 \\
    \hline
    28kroA100-4x7 & 161087.0 & 173691.0 & 0.45  & -     & -     & -     & 134129.0 & 134532.3 & 0.00 \\
    \hline
    2lin105-2x1 & 300290.7 & 319749.4 & 1.60  & 920629.6 & 1035690.4 & 0.12  & 152729.7 & 156396.6 & 0.00 \\
    \hline
    30kroB100-5x6 & 216499.5 & 227926.8 & 0.62  & -     & -     & -     & 198976.7 & 199205.7 & 0.00 \\
    \hline
    35kroB100-5x5 & 166362.5 & 179525.2 & 0.92  & -     & -     & -     & 129122.6 & 129832.3 & 0.00 \\
    \hline
    36eil101-6x6 & 4752.6 & 5226.1 & 0.62  & -     & -     & -     & 3850.7 & 3852.1 & 0.00 \\
    \hline
    42rat99-6x7 & 9706.4 & 10068.5 & 0.92  & -     & -     & -     & 8902.5 & 8906.0 & 0.00 \\
    \hline
    4berlin52-2x2 & 37576.7 & 43289.4 & 0.25  & 58055.8 & 65276.7 & 0.05  & 23287.9 & 23395.9 & 0.00 \\
    \hline
    4eil51-2x2 & 2691.4 & 2870.8 & 0.25  & 2866.8 & 3200.3 & 0.05  & 1898.5 & 1915.6 & 0.00 \\
    \hline
    4eil76-2x2 & 4312.7 & 4800.2 & 0.53  & 5736.3 & 6451.4 & 0.08  & 2948.8 & 2974.6 & 0.00 \\
    \hline
    4pr76-2x2 & 747062.3 & 821661.9 & 0.53  & 1227486.9 & 1437245.4 & 0.08  & 442693.0 & 445997.3 & 0.00 \\
    \hline
    6berlin52-2x3 & 40772.6 & 43360.3 & 0.38  & -     & -     & -     & 32130.8 & 32295.7 & 0.00 \\
    \hline
    6pr76-2x3 & 747967.5 & 822531.1 & 0.38  & 1049553.9 & 1226278.5 & 0.07  & 648884.9 & 656228.6 & 0.00 \\
    \hline
    6st70-2x3 & 4287.0 & 4631.6 & 0.32  & 5354.6 & 6458.5 & 0.07  & 3476.7 & 3503.5 & 0.00 \\
    \hline
    8berlin52-2x4 & 35300.7 & 40698.9 & 0.32  & -     & -     & -     & 26854.4 & 26969.8 & 0.00 \\
    \hline
    9eil101-3x3 & 4397.4 & 4893.6 & 0.38  & 6491.1 & 8198.1 & 0.08  & 3135.4 & 3154.3 & 0.00 \\
    \hline
    9eil51-3x3 & 2197.0 & 2293.3 & 0.38  & 2841.6 & 3382.1 & 0.08  & 1912.8 & 1921.3 & 0.00 \\
    \hline
    9eil76-3x3 & 3761.3 & 3880.8 & 0.32  & 4945.0 & 5768.9 & 0.08  & 2938.4 & 2956.6 & 0.00 \\
    \hline
    9pr76-3x3 & 738481.6 & 778897.8 & 0.32  & 1012732.3 & 1227532.7 & 0.08  & 554995.8 & 560230.3 & 0.00 \\
    \hline
    \end{tabular}%
  \label{tab:ResultsType6}%
\end{table*}%